\documentstyle[12pt]{article}
\begin{document}
\begin{center} \huge {HIGGS $\rightarrow $ TWO PHOTON DECAY:}
\end{center}
\begin{center}
      \huge {QCD radiative correction.}
\end{center}
\vspace{0.2cm}
\begin{center} \bf {K. Melnikov$^{(1)}$ and O. Yakovlev$^{(2)}$}
\end{center}
\vspace{0.1cm}
\begin{center} $^{(1)}$ Novosibirsk State University , 630090
Russia, Novosibirsk \\
 $^{(2)}$ Budker Institute for Nuclear Physics , 630090
Russia, Novosibirsk \\
\end{center}
\vspace{0.2cm}
\begin{center} \bf {Abstract}
\end{center}
QCD radiative correction to Higgs $\rightarrow$ two photons decay
rate is calculated. Below the threshold we found negligible
correction, thus supporting results obtained earlier by Djouadi
et all [7]. Above the threshold radiative correction appears to be large
for both real and imaginary part of H $\gamma \gamma$ vertex.
This leads to radiative correction for $\Gamma $(H $\rightarrow \gamma \gamma
$)
to be of order 20--100 percents at $m_{t}=150 GeV$.
 Possible applications of our results
for Higgs search at Next Linear Colliders (NLC) are briefly discussed.

\begin{center} {\bf 1. Introduction}
\end{center}
\par
The problem of Higgs boson hunting remains to be the most important
problem of contemporary high-energy physics. Different ideas and suggestions
 were put forward in this direction.
\par
Recently, the two-photon decay mode of Higgs-boson attracted much
attention of both theorists and experimentalists. This interest
is based on two different stories:
\begin{itemize}
\item firstly, this decay chanel, having obviously small rates (
 Br(H $ \rightarrow \gamma \gamma ) \sim O(10^{-3} $) for $ m_H \approx $ 150
 GeV ) provides us attractive possibility to discover and study
 Higgs boson in the intermediate mass range  at hadron colliders
 such as LHC and SSC [1].
\item secondly, two photon production of heavy Higgs boson via $
\gamma \gamma \to $ H $ \to X \quad (X= ZZ, t \bar t ) $ is quite interesting
and promising. One can hope to study the contribution of nonstandard
ultraheavy particle to $ \Gamma (H \to \gamma \gamma ) $, probe different
 anomalous interactions [2-4] using this very reaction. It was also
 suggested [5] to use $ \gamma \gamma \to H \to t \bar t $ to
 study Yukawa Higgs-top coupling.
\end{itemize}
\par
The important point for all these discussions is the two-photon width
of the Higgs boson, which was calculated to leading order in ref.[6].
In this letter we report the calculation of QCD radiative correction to
$ H \to \gamma \gamma $ decay channel. This radiative correction seems to
be the largest within the Standard Model and one must know its value
in order to exploit both two -- photon decay and production of Higgs
 boson.
 \par
 It must be noted that similar calculation already exists in the
 literature [7]. However, the authors of ref. [7] restricted
 themselves to the calculation of QCD radiative correction
 for Higgs boson lighter then $ 2 m_t $ only.
 \par
 In our work we tried to improve the situation and have solved the
 problem for the whole range of Higgs masses.

\begin{center}{\bf 2. Method of calculation }
\end{center}
\par
As it is well-known [6], Higgs -- two photon interaction can be
described by the effective Lagrangian:
\begin{equation}
  {\cal L} ={\frac {\alpha F }{4 \pi }} (\sqrt {2} G_{F})^{\frac {1}{2}}
   F_{\mu \nu } F^{\mu \nu } H
\end{equation}
\par
Here $\alpha $ is the fine structure constant,
 $ G_F $ is Fermi coupling constant
 while $ F_{ \mu \nu } $ and $ H $ stand for photon and Higgs fields
respectively.
$ F $ is Higgs -- two photon formfactor, which reads:
\begin{equation}
F=\sum_{i} N_{c_{i}} {Q_{i}}^2 f_i(\frac{{m_H}^2}{{m_i}^2})
\end{equation}
\par
Here $N_{c_i}$ is a number of colors, $ Q_i $ is  charge of the particle
and summation is performed over all particles which contribute
to the loop. $ {f_{i}}^{0} $ were computed in ref.[5]:
\begin{displaymath}
{f_t}^{(0)}= - 2 {\beta}_{t} ((1- {\beta}_{t}) \cdot x^2+1)
\end{displaymath}
\begin{displaymath}
{f_W}^{(0)}=  2 +3 {\beta}_{W} +3{\beta} _{W} \cdot (2- {\beta}_{W} ) \cdot x^2
\end{displaymath}
Here
\begin{displaymath}
x  =  arctg(\frac {1}{\sqrt {\beta_{i} -1} }) \qquad  \beta_{i} > 1
\end{displaymath}
\begin{displaymath}
x  =  \frac {1}{2} (i \cdot log(\frac {1 +\sqrt {1-\beta_{i}}}
{1 -\sqrt {1-\beta_{i}}})+ \pi ) \qquad  \beta_{i} < 1
\end{displaymath}
\begin{displaymath}
{\beta }_{i} =\frac {4{m_i}^2}{{m_H}^2}
\end{displaymath}
 It is sufficient to consider only top's and $ W $ contribution;
light quarks interaction with Higgs are negligible.
\par
As W does not interact with gluons, we need top's contribution only
to compute QCD radiative correction. Generic diagramms are
shown in Fig.2.
\par
Analyzing the structure of the Feynman graphs in Figs. 1 and 2, we
can write a simple formulae, which allows us to compute the contribution
of a given graph directly to the formfactor:
\begin{equation}
f_{(\alpha )}= \frac {d_{\mu \nu } {T_{(\alpha )}}^{\mu \nu }(k_1,k_2)}
{d_{\mu \nu } d^{\mu \nu }}
\end{equation}
\par
Here $ d^{\mu \nu }$  stand for
$ (g^{\mu \nu } k_1 \cdot k_2 - {k_1}^{\nu }{k_2}^{\mu }) $
 and $ {T_{(\alpha )}}^{\mu \nu }(k_1,k_2) $  is the amplitude which
 corresponds to the graph labelled $ \alpha $ ( One must keep in mind
 that the contribution of a single graph is not gauge -- invariant
 itself, but applying Eg. (3) we can work with its  gauge -- invariant
 piece only.)
 \par
 Obviously, $ H \gamma \gamma $ formfactor ( we take only top quark
 contribution into account ) is analytic function with a cut over
 real axis from $ 4 {m_t}^2 $ to infinity.
 Thus, $ f_t (s) $ obeys dispersion relation, which reads
\begin{equation}
f_t (s)=\frac {1}{\pi } \int \limits_{4 {m_t}^2 }^{\infty}
  \frac {Im f_t(s')}{s'-s-i \varepsilon } ds'
\end{equation}
\par
For our mind, using dispersion relation (Eq.(4)) is  the best way
to calculate the contribution of graphs in Fig.2
to the formfactor. Thus, our work proceeds in two steps:
firstly, we compute $ Im f_t (s) $ analytically and then,using
Eq.(4) we perform numerical integration and obtain the answer valid
for arbitrary Higgs mass.
\par
In order to compute the contribution of graphs in Fig.2 to
$ Im f_t (s) $ we must cut them in a well-known manner, facing the
graphs with two or three particles in the intermediate state (see Fig.3
for illustration ).The latter are Born graphs, while the former
include one--loop proper subgraphs such as $ \gamma t \bar t $ vertex,
top's self-energy and $ H t \bar t $ vertex. Those subgraphs
diverge and hence must be renormalized. On this way we used on--shell
scheme for quark propagator renormalization and, as is well-established in
QED, we performed subtraction for $ \gamma t \bar t $ in zero momentum
transfer thus satisfying Ward identities. As for $ H t \bar t $
vertex, the one--loop counter--term is fixed by top's mass and wave
function renormalization due to the fact that it is Higgs--fermion
interaction which gives the mass to the fermion ( for exhaustive
discussion see ref.[8] )
Analitical calculations were performerd, by Reduce 33 , on the base
of the package, written by us.

\begin{center} \bf  {3. Results}
\end{center}
\par
Following the way, outlined in the previous section, we obtained
$ f_t (s) $ and $ F(s) $ on two-loop level. In order to present our results
in the convenient form, we will need some notations. Let us write
two-loop QCD corrected formfactor in such a way
\begin{equation}
F(s)={F}^{(0)}(s)+ \frac {{\alpha}_{s}}{\pi} F^{1}(s)
\end{equation}
\par
$ {F}^{(0)} $ is the lowest order formfactor, defined in eq.(2).

\par
Our results for $ Re {F}^{(0)}(\beta ), Re {F}^{(1)}(\beta ) $
and $ Im {F}^{(0)}(\beta ), Im {F}^{(1)}(\beta )$ are plotted in Figs. 4,5
respectively. We used the value of the top mass to be 150 GeV,
and for W we took 80.26 GeV.
Thus, below the threshold our results coincide with
that of ref.[7]. Above the threshold radiative correction to
$ F^{(0)} (\beta ) $ is large everywhere ( $\approx $ 30--40 $ \% $ ),thus
reminding us
about similar situation in $ H \to b \bar b $ , $ H \to V \gamma $
studies.
\par
It is straightforward then to compute QCD radiative correction to Higgs
$ \rightarrow $ two photons decay rate.   Corresponding curves
are plotted in Fig.6. Below threshold, radiative correction is
negligible. Above the threshold its value is about 20--100 percents at
$ m_{t}=150 GeV $.
We want to make some comments for such enormously large
radiative correction. This large radiative correction arises
for next  two reasons:
\begin{enumerate}
\item large  QCD radiative corrections to purely top's contribution
to formfactor;
\item well-known compensation of $ W $ and $ t $ contribution to
the formfactor. Roughly, this compensation occurs for $ Im F^{(0)} $
for $ m_H $ above 600 GeV, while for $ Re F^{(0)} $ this phenomenon
takes place for $ m_H $ below 600 GeV.
\end{enumerate}
\par
Due to this facts, the first (negative) peak ( $ m_H \approx 460 GeV $)
 in Fig.6 is controlled by the correction to the imaginary part
 of the formfactor while the second  ( $ m_H \approx 660 GeV $)
 is governed by the correction to the real part of the formfactor.

\begin{center} {\bf 4.Discussion.}
\end{center}
\par
We have computed QCD radiative correction to $ H \to \gamma \gamma $
decay rate. As mentioned in the introduction, our results can be used
for studying the possibility of Higgs search at Next Linear Colliders.
Below two tops threshold QCD correction is negligible ( about 1 $ \% $).
Above the threshold, radiative correction is large enough (20--100 $ \% $)
at $m_{t}=150 GeV$ and more at $ m_{t}=150-200 GeV $.
\par
For example, for $ \gamma \gamma \to H \to Z Z $ [2], the total cross
section is proportional to $ \Gamma ( H \to \gamma \gamma) $. Thus
$ \Delta \sigma (\gamma \gamma \to H \to ZZ ) $ is equal to
$ \Delta \Gamma (H \to \gamma \gamma ) $ . ( Here by $\Delta $ we
mean relative radiative correction of the corresponding quantity ).
As this correction for $  m_H  \approx 600 GeV $ appears to be about
$ 100 \% $,
this fact must be taken into account while discussing the validity of this
channel for probing Higgs sector.
\par
Another consequence of our result is connected with proposed study [5]
of the Yukawa coupling in $ \gamma \gamma \to H \to t \bar t $. (In this
case the effect arise from the interference of $ \gamma \gamma \to H \to t \bar
t $
with the Born amplitude $ \gamma \gamma \to t \bar t $  in the
resonance region ( $\sqrt {s_{\gamma \gamma }} \approx m_H $ ).
It is imaginary part of $ H \gamma \gamma $ vertex that contribute to
this interference. Due to the fact that $ W $ and $ t $ contributions
interfere distructivly, the change in the $ Im f_t (\tau ) $ due
to QCD correction will roughly double the signal in comparison
with the results of ref.[5] for $ m_H \geq $ 700 GeV.
However, for such a heavy Higgs, the signal is so hardly observable
that this result must not be taken too seriously.
\par
It was also suggested recently [9] to study $ \gamma \gamma \to HH $
in order to probe anomalous triple Higgs vertex. We want to note here
that as QCD correction to $ \gamma \gamma \to H $ appears to be large,
this must be taken into account for proper discussion of
$ \gamma \gamma \to H H $ cross-section sensitivity for triple
Higgs anomalous coupling. It seems, that spoiling interference
of $ W $ and top contribution will inspire better manifestation of
possible anomalies.
\begin{center} { \bf Acknowledgement:}
\end{center}
 The authors are very gratefull to Profs. V.S.Fadin and I.F.Ginzburg
 for numeruos discussions and support during the work on the problem.
\begin{center} { \bf Referencies}
\end{center}
\begin{enumerate}
\item
  \begin{enumerate}
  \item Higgs Working Group, D. Froidevaux, Z. Kunszt, W.J.Stirling,
        report ECFA LHC Workshop (CERN/Aachen, October, 1990).
  \item M. Schneegans, LAPP--ANNECY preprint, LAPP-EXP-91.
  \end{enumerate}
\item I.F. Ginzburg, IM-28(182) preprint, 1990.
\item G. Belanger, F. Boudjema, preprint EVSLAPP-LPN-TH-80, 1992.
\item H. Konig, Phys. Rev. D , 45 (1992), 1575.
\item E.E. Boos et. all, preprint MPI-PH/92-48.
\item J.Ellis, M.K.Gaillard, D.Nanopoulos. Nucl. Phys., B106, 292, 1976.\\
      A.I. Vainstein et. all, $ \quad $ Sov.J.Nucl.Phys.,1979,v.30,p.1368.
\item A.Djouadi et all, Phys.Lett.B, 257 (1991), 187.
\item M. Drees, K. Hikasa, Phys.Lett.B 240 (1990), 455.
\item G.V. Jikia, IHEP 92-91 preprint.
\end{enumerate}
\end{document}